\documentstyle[preprint,psfig,aps,12pt,amssymb]{revtex}\tighten
\begin{document}
\title{
\rightline{\small{{\em J. Math. Phys.\/}{\bf 35}, 4184 (1994)}}
Null cone evolution of axisymmetric vacuum spacetimes}
\author{R. G\'{o}mez, P. Papadopoulos and J. Winicour }
\address{Department of Physics and Astronomy,
 	University of Pittsburgh,
 	Pittsburgh, PA 15260}
\maketitle

\begin{abstract}
We present the details of an algorithm for the global evolution of
asymptotically flat, axisymmetric spacetimes, based upon a
characteristic initial value formulation using null cones as evolution
hypersurfaces. We identify a new static solution of the vacuum field
equations which provides an important test bed for characteristic
evolution codes. We also show how linearized solutions of the Bondi
equations can be generated by solutions of the scalar wave equation,
thus providing a complete set of test beds in the weak field regime.
These tools are used to establish that the algorithm is second order
accurate and stable, subject to a Courant-Friedrichs-Lewy condition.
In addition, the numerical versions of the Bondi mass and news
function, calculated at scri on a compactified grid, are shown to
satisfy the Bondi mass loss equation to second order accuracy. This
verifies that numerical evolution preserves the Bianchi identities.
Results of numerical evolution confirm the theorem of Christodoulou and
Klainerman that in vacuum, weak initial data evolve to a flat spacetime. 
For the class of asymptotically flat, axisymmetric  vacuum spacetimes, for
which no nonsingular analytic solutions are known, the algorithm
provides highly accurate solutions throughout the regime in which
neither caustics nor horizons form.
\end{abstract}


\section{Introduction}
The physical basis of a new algorithm for the evolution of
spacetimes has been proposed.~\cite{IWW} This algorithm is based upon
the characteristic initial value problem for general relativity, using
light cones as the evolution hypersurfaces, rather than the spacelike
foliation used in traditional approaches based upon the Cauchy
problem.  The intimate use of characteristics has particular advantages
for the description of gravitational radiation and black hole
formation.~\cite{EKG,waveforms,south}. The first attempt to carry out
numerical evolution based upon this algorithm was only successful in a
region outside some sufficiently large worldtube. Near the vertex of
the null cone, instabilities arose which destroyed the accuracy of the
code. The underlying cause of this instability was too complicated to
analyze in the context of general relativity, especially since the
numerical analysis of the characteristic initial value problem had not
yet been carried out even for the simplest linear axisymmetric systems.

This warranted an investigation of the basic computational properties of
the evolution of the flat space scalar wave equation using a null cone
initial value formulation~\cite{jcomp}. It was found that near the
vertex of the cone the Courant-Friedrichs-Lewy (CFL) condition places a
stricter limit on the time step than for the case of Cauchy evolution.
This insight made possible the development of an extremely efficient
marching algorithm for evolving the data on the initial cone by
stepping it out from the vertex of the next cone to null infinity
(scri) along each angular ray direction. This marching algorithm is
based upon a simple identity satisfied by the values of the scalar
field at the corners of a parallelogram formed by four null rays. The
result was a stable, calibrated, second order accurate global algorithm
on a compactified grid. Furthermore, scri played the role of a perfect
absorbing boundary so that no radiation was reflected back into the
system. This algorithm offers a powerful new approach to generic wave
type systems.  The basic idea is applicable to any of the
hyperbolic systems occurring in physics.

In this paper, we apply the algorithm to the evolution of
asymptotically flat, (twist-free) axisymmetric spacetimes. In a Bondi
null coordinate system the metric takes the form~\cite{bondi}
\begin{eqnarray}
ds^2 & = & ({V \over r}  e ^{2 \beta} 
     - U^2 r^2 e^{2 \gamma}) du^2 
     + 2 e^{2 \beta} du dr 
     + \  2 U r^2 e^{2 \gamma} du d\theta \nonumber \\
     &   & - r^2 (e^{2 \gamma} d\theta^2
        + e^{-2\gamma} \sin^2 \theta d\phi^2). 
\label{eq:bmetric}
\end{eqnarray}
The vacuum field equations then decompose into the three hypersurface 
equations
\begin{equation}
\beta_{,r} = \frac{1}{2} r\: (\gamma_{,r})^2  \label{eq:beta}
\end{equation}
\begin{equation}
   [r^4 \, e^{2(\gamma-\beta)} U_{,r}]_{,r} =
   2 r^2 [ r^2  (\frac{\beta}{r^{2}})_{,r\theta} 
   - \frac{(\sin^2 \theta \,\gamma)_{,r\theta}}{\sin^{2}\theta} 
   + 2\,\gamma_{,r} \, \gamma_{,\theta}) ]  \label{eq:U}
\end{equation}
\begin{eqnarray}
   V_{,r} & = &-\frac{1}{4} r^4 e^{2(\gamma-\beta)}(U_{,r})^2 
   + \frac{(r^4 \sin\theta U )_{,r\theta}}{2 r^2 \sin\theta}
   \nonumber \\
   & & +e^{2 (\beta - \gamma)}[ 
       1 - \frac{(\sin \theta \beta_{,\theta})_{,\theta}}{\sin\theta} 
       + \gamma _{,\theta\theta} + 3 \cot\theta \gamma_{,\theta} 
       - (\beta_{,\theta})^2 - 2 \gamma_{,\theta} (\gamma_{,\theta}
       -\beta_{,\theta}) ]            \label{eq:V}
\end{eqnarray}
and one evolution equation
\begin{eqnarray}
4 r ( r \gamma)_{,ur}  & = &  [ 2 r \gamma_{,r} V  
- r^{2} ( 2 \gamma_{,\theta}\,U  
          + \sin \theta ({U \over {\sin \theta}})_{,\theta} )]_{,r} 
-2 r^{2} \frac{(\gamma_{,r} U \sin\theta)_{,\theta}}{\sin\theta} 
\nonumber \\
& & + \frac{1}{2} r^{4} e^{2(\gamma-\beta)} (U_{,r})^{2} 
+ 2 e^{2(\beta - \gamma)} [ (\beta_{,\theta})^2 
     + \sin \theta ({\beta_{,\theta} \over {\sin \theta}})_{,\theta} ] .  
\label{eq:gammaev}
\end{eqnarray}
The initial data consists of $\gamma$, which is unconstrained except by
smoothness conditions. Because $\gamma$ represents a spin-2 field, it
must be $O(\sin^2 \theta)$ near the axis and consist of $l\ge 2$ spin-2
multipoles.

Here we are interested in the global application of this system when
the null hypersurfaces are null cones, although the approach also goes
through without major change if the null hypersurfaces emanate from a
finite world tube.  We require that the null cones have nonsingular
vertices which trace out a geodesic worldline $r=0$. For the quadrupole
terms, this implies the boundary conditions $\gamma=O(r^2)$,
$\beta=O(r^4)$, $U=O(r)$ and $V=r+O(r^3)$.  For higher multipoles, the
smoothness conditions can be worked out by referring back to local
Minkowski coordinates \cite{twistor1}. As a consequence, $O({r^n})$ terms
in $\gamma$ can contain multipoles with $2\le l \le n$. Any satisfactory
computational algorithm must meet the challenge of preserving these
smoothness conditions. 

In Sec.~\ref{sec:linear}, we analyze the linearized version of these
equations and show how their solutions may be obtained locally from
solutions of the scalar wave equation. This provides an important means
of constructing linearized solutions in a null cone gauge for the
purpose of code calibration. It also reveals essential changes in the
grid structure necessary in adapting the null parallelogram algorithm
for the wave equation to linearized gravity.

This also solves the major computational problems for the nonlinear
case.  In Sec.~\ref{sec:nonlinear}, we show how the linearized
algorithm can be extended to this case. In Sec.~\ref{sec:fde}, we
discuss the major finite difference techniques necessary for second
order accuracy.  In Sec.~\ref{sec:tests}, we present a study of the
stability and accuracy of an evolution code based upon this algorithm.
New exact and linearized solutions are introduced to establish second
order convergence. In addition, a global check on accuracy is carried
out using Bondi's formula relating mass loss to the time integral
of the square of the news function.

\section{The Linearized Bondi Equations} \label{sec:linear}
In the linearized limit  $\beta = 0$ and $V = r$.
The equations (\ref{eq:beta})-(\ref{eq:gammaev}) reduce to one hypersurface equation
and one evolution equation for the surviving field variables $\gamma$
and $U$, 
\begin{equation}
(r^4  U_{,r})_{,r} = -  
\frac{2 r^2 (\sin^2 \theta \,\gamma)_{,r\theta}}{\sin^{2}\theta}  
\label{eq:lu}
\end{equation}
\begin{equation}
4 r ( r \gamma)_{,ur} = [ 2 r^{2} \gamma_{,r}  
- r^{2}  \sin \theta ({U \over {\sin \theta}})_{,\theta}]_{,r}. 
\label{eq:lgammaev}
\end{equation}

Physical considerations suggest that these equations be related to the
wave equation. If this relationship were sufficiently simple, then the
scalar wave algorithm could be used as a guide in formulating an
algorithm for evolving $\gamma$.   A scheme for generating solutions to
the linearized Bondi equations in terms of solutions to the wave
equation has been presented previously~\cite{bondi}. However, in that
scheme, the relationship of the scalar wave to $\gamma$ is non-local in
the angular directions and is not useful for this purpose.

We now formulate an alternative scheme in terms of
spin-weight 0 quantities $\alpha$ and $Z$, related to
$\gamma$ (spin-weight 2) and $U$ (spin-weight 1) by~\cite{conserved}
\begin{equation}
   \gamma = \eth ^2 \alpha = \sin \theta \partial_{\theta}
            ({1 \over{\sin \theta}} \partial_{\theta} \alpha)
    \label{eq:edthsq}
\end{equation}
\begin{equation}
      U=\eth Z=\partial_{\theta} Z.
\end{equation}
Then the linearized equations are equivalent to
\begin{equation}
 (r^4  Z_{,r})_{,r}  =  -2 r^2(2-L^2)\alpha_{,r} ,
  \label{eq;hz} 
\end{equation}
 and
\begin{equation}
  E:=2( r \alpha)_{,ur} -r^{-1}(r^2 \alpha_{,r}-r^2 Z/2)_{,r}=0,
 \label{eq:E}
\end{equation}
where $L^2= -(1/\sin\theta)\partial_{\theta}(\sin\theta\partial_{\theta})$ 
is the $\theta$-part of the angular momentum operator. Now let $\Phi$ be a
solution of the flat space scalar wave equation,
\begin{equation} 
   r \Box\, \Phi = 2(r\Phi)_{,ur}- (r\Phi)_{,rr} +r^{-1}L^2 \Phi =0,
    \label{eq:wave}
\end{equation}
and set
\begin{equation} 
      r^2 \alpha_{,r}=(r^2\Phi)_{,r} \label{eq:alpha} 
\end{equation}
and
\begin{equation}
       r^2 Z_{,r}= 2(L^2 -2)\Phi.  \label{eq:Z}
\end{equation}
Then
\begin{equation}
  E= r\Box \, \Phi +2(\Phi +\alpha)_{,u}-2r^{-2}(r^2 \Phi)_{,r} +Z,
\end{equation}
and
\begin{equation}
  E_{,r}= r^{-2}(r^3 \Box \, \Phi )_{,r}
\end{equation}
Equation (\ref{eq;hz}) is satisfied as a result of (\ref{eq:alpha}) and
(\ref{eq:Z})and the wave equation (\ref{eq:wave}) implies $E_{,r}=0$.
If $\Phi$ is smooth and $O(r^2)$ at the origin, this implies $E=0$, so
that the linearized equations are satisfied. The condition that
$\Phi=O(r^2)$ eliminates fields with only monopole and dipole
dependence so that it does not restrict the generality of the
spin-weight 2 function $\gamma$ obtained through (\ref{eq:edthsq}).

Thus for any linearized axisymmetric gravitational wave in the null cone gauge,
$\gamma$ may be related to a scalar wave by (\ref{eq:edthsq}) and
(\ref{eq:alpha}). The CFL condition for convergence of a finite
difference algorithm requires that the numerical domain of dependence
be larger than the physical domain of dependence. Because the
relationship between $\Phi$ and $\gamma$ is local with respect to the
null rays on the cone, their domains of dependence coincide. This
suggests that a stable and convergent evolution algorithm for the
gravitational field may be modeled upon the scalar wave algorithm. This
turns out to be the case although some subtle distinctions arise, as
described below.

\begin{figure}
\centerline{\psfig{figure=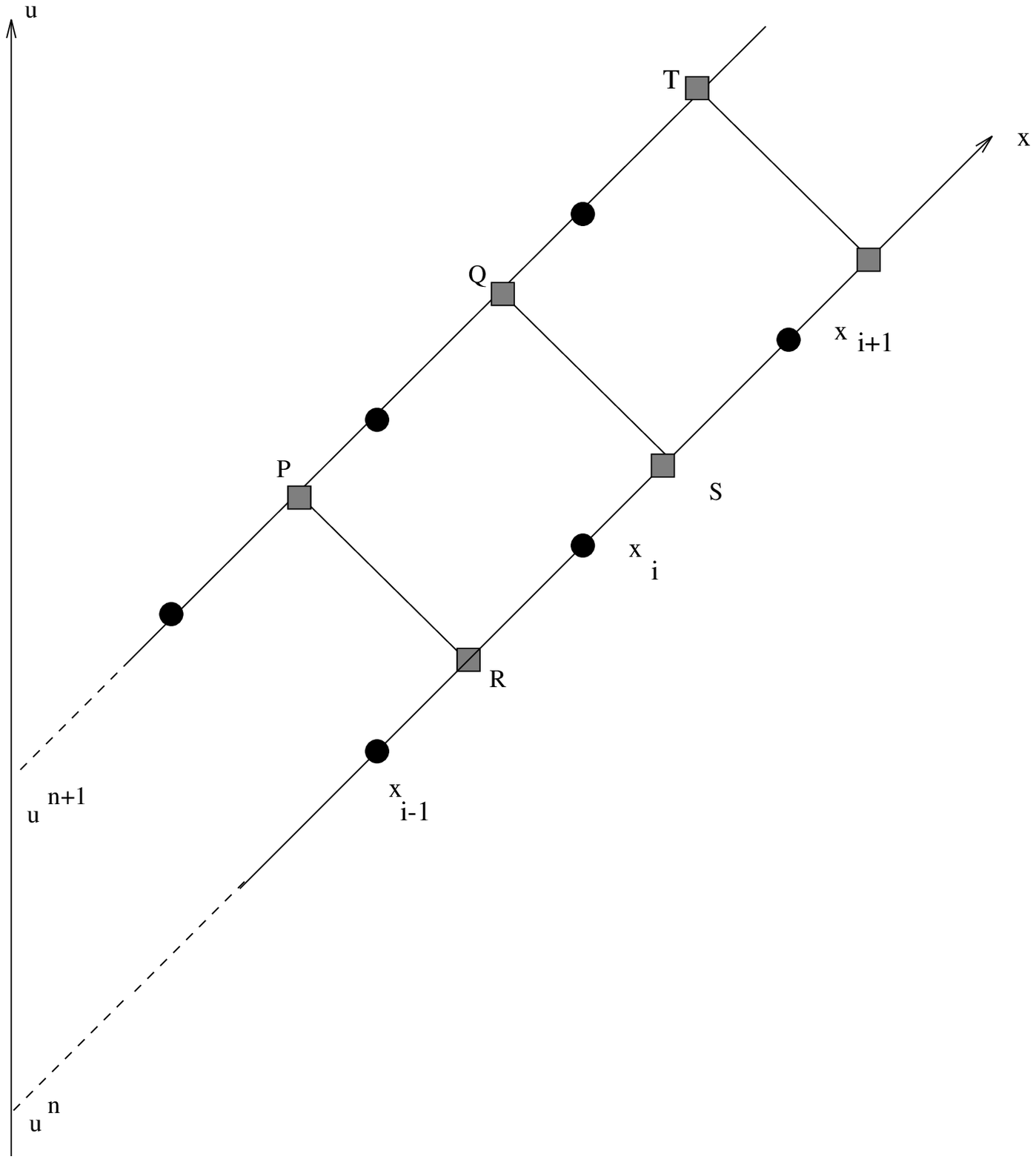,height=7.765in,width=6.0in}}
\caption{Line segments drawn at forty-five degrees represent radial
characteristics. Their intersection defines the fundamental null
parallelogram PQRS shown superimposed upon the computational cell, which consists of the points marked by circles and their nearest neighbors in
the angular direction.}
\label{fig:cell}
\end{figure}

An evolution algorithm for scalar waves has been formulated in terms of
an integral identity for the values of the field at the corners of a
null parallelogram lying on the $(u,r)$ plane~\cite{jcomp}. The wave
equation with source, $\Box \Phi =S$, can be reexpressed in the form
\begin{equation}
       \Box\, ^{(2)} \psi = -{L^2 \psi \over r^2}+rS, \label{eq:sbox}
\end{equation}
where $\psi=r\Phi$ and $\Box\, ^{(2)}$ is the 2-dimensional wave
operator intrinsic to the $(u,r)$ plane. Integration over the null
parallelogram as depicted in Fig.~\ref{fig:cell} then leads to
the integral equation
\begin{equation}
     \psi_Q = \psi_P + \psi_S - \psi_R
     +{1 \over 2} \int_A du dr [-{L^2 \psi \over r^2}+rS],
  \label{eq:snp}
\end{equation}
where $P$, $Q$, $R$ and $S$ are the corners and $A$ the area of the
parallelogram.

This identity gives rise to an explicit marching algorithm for
evolution. Let the null parallelogram span null cones at adjacent grid
values $u_0$ and $u_0+\Delta u$, as shown in
Fig.~\ref{fig:cell}, for some $\theta$ and $\phi$. If $\psi$
has been determined on the entire $u_0$ cone and on the $u_0+\Delta u$
cone radially outward from the origin to the point $P$, then
(\ref{eq:snp}) determines $\psi$ at the next point $Q$ in terms of an
integral over $A$. This procedure can then be repeated to determine
$\psi$ at the next radially outward point $T$ in
Fig.~\ref{fig:cell}.  After completing this radial march to
scri, the field $\psi$ is then evaluated on the next null cone at
$u_0+2\Delta u$, beginning at the vertex where smoothness gives the
start up condition that $\psi=0$. The resulting evolution algorithm is
a 2-level scheme which reflects, in a natural way, the distinction
between characteristic and Cauchy evolution, i.e. that the time
derivative of the field is not part of the characteristic initial
data.

The CFL condition on the numerical domain of dependence is a
necessary condition for convergence of a numerical algorithm. For the
grid point at $(u,r,\theta)$, there are three critical grid points,
$(u-\Delta u,r+\Delta r,\theta)$ and $(u-\Delta u,r-\Delta r,\theta
\pm \Delta \theta)$, which must lie inside its past physical
light cone. These gives rise to the inequalities $\Delta u < 2\Delta r$
and $\Delta u < -\Delta r+(\Delta r^2 +r^2 \Delta \theta^2)^{1/2}$. At
large $r$, the second inequality becomes  $\Delta u < r\Delta \theta$
and the limitations on the time step are essentially the same as for a
Cauchy evolution algorithm. However, near the vertex of the cone, the
second inequality gives a stricter condition
\begin{equation}
      \Delta u < K\Delta r \Delta \theta^2, \label{eq:cfl}
\end{equation} 
where $K$ is a number of order $1$ whose exact value depends upon the
start up details at the vertex. For the scalar wave equation, these
stability limits were confirmed by numerical experimentation and it was
found~\cite{jcomp} that $K\approx 4$.

The linearized gravitational evolution equation (\ref{eq:lgammaev}) can be recast into a form similar to (\ref{eq:sbox}),
\begin{equation}
       \Box\, ^{(2)} \psi = {\cal H}, \label{eq:lbox}
\end{equation}
where now $\psi=r \gamma$ and ${\cal H}$ only contains hypersurface
terms.  This allows use of the null parallelogram algorithm to evolve
$\gamma$ by the same marching scheme as in the scalar case. The
additional feature here is that $U$ must be simultaneously marched out
the null cone using the hypersurface equation (\ref{eq:lu}). For the
scalar wave equation (\ref{eq:sbox}), the angular momentum
barrier is represented by the term $L^2 \psi / r^2$, which is
determined from $\psi$ by a double angular derivative. In the
linearized gravitational evolution equation (\ref{eq:lgammaev}), the
analogous term is $[r^{2} \sin \theta ({U /\sin
\theta})_{,\theta}]_{,r}$, which is determined from $U$ by a single
angular derivative. In turn, the hypersurface equation (\ref{eq:lu})
relates $U$ to $\gamma$ by a single angular derivative. Physically,
this has the same net effect of producing an angular momentum barrier
depending upon the second angular derivative, as in the scalar case.
However, the nontrivial mathematical distinction between the two cases
leads to nontrivial difference in their natural grid structures for a
numerical algorithm. In particular, the grid for $U$ must be staggered
between the grid points for $\gamma$. These and other details of the
marching algorithm will be given in Sec.~\ref{sec:fde}, where we
discuss the  nonlinear case.

The use of scalar waves to generate solutions of the linearized Bondi
equations provides an important tool for testing evolution algorithms
in the linear regime. Monopole solutions may be represented in the form
$\Phi=[F(u+2r)-F(u)]/r$ and axisymmetric multipoles may be built up by
applying the $z$-translation operator
\begin{equation}
       \partial_z = \cos \theta(\partial_r -\partial_u)
                   -r^{-1} \sin \theta \partial_{\theta}
\end{equation}
to these solutions. Then $\gamma$ may be obtained via
(\ref{eq:edthsq}). For the calibration measurements in
Sec.~\ref{sec:tests}, we use the solutions
\begin{equation}
       \Phi=(\partial_z)^{\ell}[u(u+2r)]^{-1}, \label{eq:ltest}
\end{equation}
obtained by applying $(\partial_z)^{\ell}$ to the fundamental Lorentz
invariant solution $1/(x^{\alpha}x_{\alpha})$. This solution is well
behaved above the singular light cone $u=0$.

\section{The Nonlinear Algorithm} \label{sec:nonlinear}
The nonlinear evolution equation (\ref{eq:gammaev}) can also be recast
in the form of (\ref{eq:lbox}) in terms of $\psi=r \gamma$, for an
appropriate choice of 2-dimensional wave operator $\Box\, ^{(2)}$. In
this case, the $(u,r)$ submanifold is not flat so that it would not be
appropriate to base $\Box\,^{(2)}$ upon a flat metric. Indeed, such a
choice would lead to an improper domain of dependence and could not be
used as the basis for a stable algorithm. It would seem more natural to
use the $\Box\,^{(2)}$ operator of the metric induced in the $(u,r)$
submanifold by the 4-dimensional metric (\ref{eq:bmetric}). Here we
pursue an alternative choice based upon the line element
\begin{equation}
  d\sigma^2=2l_{(\mu}n_{\nu )}=e^{2 \beta} du[\frac{V}{r} du +2 dr],  
   \label{eq:2metric}   
\end{equation}
where $l_\mu =u_{,\mu}$ is the normal to the outgoing null cones and
$n_\mu$ is the other null vector normal to the spheres of constant
$r$.  Although this choice is not unique and other possibilities
deserve exploration, it leads to the simplest ${\cal H}$-terms when
reexpressing (\ref{eq:gammaev}) in the form (\ref{eq:lbox}). Because
the domain of dependence of $d\sigma^2$ contains the domain of
dependence of the induced metric of the $(u,r)$ submanifold, this
approach does not lead to convergence problems associated with the CFL condition.

The wave operator associated with (\ref{eq:2metric}) is 
\begin{equation}
    \Box\,^{(2)}\psi=e^{-2 \beta}[2\psi_{,ru}-(\frac{V}{r} \psi_{,r})_{,r}]
   \label{eq:2box}   
\end{equation}
and the nonlinear evolution equation (\ref{eq:gammaev}) becomes
\begin{equation}
        \Box\, ^{(2)} \psi = e^{-2 \beta}{\cal H}, \label{eq:nlbox}       
\end{equation}
where
\begin{eqnarray}
  {\cal H} & = & -({V\over r})_{,r}\gamma -{1 \over r}
[r^2 (\gamma_{,\theta}U
    + \frac{1}{2} \sin \theta ({U \over {\sin \theta}})_{,\theta})]_{,r} 
   \nonumber \\
& &  - r \frac{(\gamma_{,r} U \sin\theta)_{,\theta}}{\sin\theta} 
   + \frac{1}{4} r^{3} e^{2(\gamma-\beta)} (U_{,r})^{2}
  \nonumber \\
& &  + {1 \over r} e^{2(\beta - \gamma)} 
[(\beta_{,\theta})^2 
+ \sin \theta ({\beta_{,\theta} \over {\sin \theta}})_{,\theta}].  
\label{eq:hterms}
\end{eqnarray}

Because all 2-dimensional wave operators are conformally flat, with
conformal weight $-2$, the surface integral of (\ref{eq:nlbox}) over a
null parallelogram gives an integral equation analogous to
(\ref{eq:snp}),
\begin{equation}
     \psi_Q = \psi_P + \psi_S - \psi_R
     +{1 \over 2} \int_A du\, dr {\cal H}.
  \label{eq:np}
\end{equation} 
This allows the evolution of $\gamma$ by the same basic marching
algorithm as described in Sec.~\ref{sec:linear} for the scalar wave
and linearized wave cases. The additional modifications are that
$\beta$, $U$, and $V$ must be simultaneously marched out the null cone
using the nonlinear hypersurface equations
(\ref{eq:beta})-(\ref{eq:V}).  Because of the  hierarchal structure of
these equations, $\gamma$, $\beta$, $U$, and $V$ may be marched in that
order without any matrix inversions or other implicit operations. The
basic computational cell and finite difference constructions are
described in the next section.

Near the origin, the metric approaches the Minkowski metric so that the
stability of the nonlinear algorithm in this region should be subject
to the same Courant limit as for the linearized equations.  Near scri,
the gravitational variables have the asymptotic form
\begin{eqnarray}
   \gamma &=& K + r^{-1} c + O(r^{-2}) \\
   \beta  &=& H + O(r^{-2}) \\ 
    U     &=& L +  O(r^{-1}) \\
        V &=& r^2 {(L \sin \theta)_{,\theta} \over {\sin \theta}} \nonumber \\
          &+&  r e^{2(H-K)} \Big[1 
             + 2 {(H_{,\theta} \sin \theta)_{,\theta} \over {\sin \theta }}
             + {(K_{,\theta} \sin^3 \theta)_{,\theta} \over {\sin^3 \theta}}
             + 4 (H_{,\theta})^2 - 4 H_{,\theta} K_{,\theta} 
             - 2(K_{,\theta})^2 \Big] \nonumber \\
          &-& 2 e^{2H} {\cal M} + O(r^{-1}) , \label{eq:Vasym}
\end{eqnarray} 
where ${\cal M}$ corresponds to Bondi's mass aspect. In a standard Bondi
frame at scri $K$, $H$, and $L$ all vanish, but not in null
cone coordinates adapted to a Minkowski frame at the origin. This
dependence can lead to drastic behavior of the $u$-coordinate at large
distances. In a numerical study of spherically symmetric,
self-gravitating scalar radiation fields~\cite{EKG}, $H\rightarrow
\infty$ as a horizon is formed and an infinite redshift develops between
central observers and observers at scri. In that case, a
consideration of the domains of dependence indicates that the step size
$\Delta u$ for stable evolution must approach zero as the horizon is
formed. The divergence of the outgoing null cone equals
$e^{-2\beta}/r$. If  $\beta \rightarrow \infty$ at a finite value of
$r$ then a caustic will in general form. When this occurs, a single
null cone coordinate system cannot cover the entire exterior region of
the spacetime.

In the more general case being considered here, it is also possible for
the $u$-direction to become spacelike at large distances, corresponding
to the coefficient of $du^2$ in (\ref{eq:bmetric}) becoming negative.
(The $u=constant$ hypersurfaces, of course, must remain null.) As
discussed in Sec.~\ref{sec:tests}, this does not affect the stability
of the algorithm.  The algorithm is also valid when the vertex of
the null cone is replaced by an inner boundary consisting of a timelike
or null worldtube, so that it may also be applied to other versions of
the characteristic initial value problem~\cite{dinv}.

\section{Finite Difference Techniques} \label{sec:fde}

The numerical grid is based upon the outgoing null cones, using the
compactified radial coordinate $x=r/(1+r)$ and the angular coordinate
$y=-\cos\theta$.  Thus points at scri are included in the grid
at $x=1$.

In order to improve numerical accuracy at the grid boundaries, the code
is written in terms of the variables
$\hat{\psi}=r\hat{\gamma}=r{\gamma}/\sin^{2}\theta$, $\beta$,
$S=(V-r)/r^2$ and $\hat{U} =U/\sin\theta$. For a pure quadrupole mode,
$\hat{\gamma}$ has constant angular dependence.

To develop a discrete evolution algorithm, we work with two sets of
spatial grid points, both of which have the constant spacing $\Delta x
=1/N_x$ and $\Delta y = 1/N_y$.  The first grid is defined by
$(u^{n},x_i,y_j)=(n\Delta u,i\Delta x,j\Delta y)$.  The second grid is
shifted (staggered) in both the $x$ and $y$ variables and is thus
defined by $(u^{n},x_{i+\frac{1}{2}},y_{j+\frac{1}{2}})=(n\Delta
u,(i+\frac{1}{2})\Delta x,(j+\frac{1}{2})\Delta y)$.  Note that the
staggered grid extends beyond the physical boundary $x=1$. This
peculiarity is successfully exploited for a smooth calculation of the
metric at scri.  The time step is variable and is limited by the
largest possible value that would satisfy the CFL condition over the
entire grid.

A staggerered grid is not necessary for the scalar wave equation but
its introduction for the gravitational case is dictated by a detailed
von Neumann stability analysis of the linearized equations. Accordingly,
$\hat{\gamma}, \beta \, \mbox{and} \, \, S$ reside on the $(x_i,y_j)$
grid while $\hat{U}$ is placed on the $(x_{i+\frac{1}{2}},y_{j+\frac{1}{2}})$ grid.
We denote values of a field $F$ at the site $(n,i,j)$ by $F^n_{ij}=
F(u^n, x_i, y_j)$. We use centered second order differences for
derivatives at points not on the edges of the grid, e.g. for
an arbitrary field $F$
\begin{equation}
{F_{,y|}}_{i,j} = {1 \over {2 \Delta y}}
                  (F_{i,j+1} - F_{i,j-1})  \label{eq:fde-y}
\end{equation}
To calculate derivatives of the field at the edges of the grid, we use
backward and forward second-order differences, e.g at the $y=\pm 1$
edges of the grid, where $j=\pm N_y$
\begin{equation}
{F_{,y|}}_{i,\pm N_y} = \mp {1 \over {2 \Delta y}}
                  (-F_{i,\pm N_y \mp 2} + 4 F_{i,\pm N_y\mp 1}
                   - 3 F_{i,\pm N_y}) \label{eq:fde-y-edge}
\end{equation}

\subsection{The Hypersurface Equations}
In terms of the numerical variables $\hat{\gamma}$, $\beta$, $S$ and
$\hat{U}$, and expressed in the coordinates $x$ and $y$ used in the code,
the hypersurface equations~(\ref{eq:beta})-(\ref{eq:V}) are
\begin{eqnarray}
\beta_{,x} &=&  {\cal H}_\beta \label{eq:xybeta} \\
({x^4 \over (1-x)^2}\, e^{2((1-y^2)\hat\gamma-\beta)}
{\hat U}_{,x})_{,x} &=& 2 {x^2 \over (1-x)^2}\: {\cal H}_U \label{eq:xyU} \\
x^2 S_{,x} + {{2 x} \over {1-x}} S  &=&  {\cal H}_S , \label{eq:xyS} 
\end{eqnarray}
where the source terms ${\cal H}_\beta$, ${\cal H}_U$ and ${\cal H}_S$ are
given by
\begin{eqnarray}
 {\cal H}_\beta &=& \frac{1}{2} x(1-x)(1-y^2)^2\: (\hat\gamma_{,x})^2 \label{eq:H-beta} \\
 {\cal H}_U &=& \beta_{,xy} - {2 \over {x(1-x)}} \beta_{,y} 
	    + 4 y \hat\gamma_{,x} 
 + (1-y^2)[2 \hat\gamma_{,x}((1-y^2)\hat\gamma_{,y} - 2 y \hat\gamma) 
             - \hat\gamma_{,xy}] \label{eq:H-U} \\
{\cal H}_S &=& 
-1 
- x y [{4 \over {1-x}}{\hat U} + x {\hat U}_{,x}]
+ {1 \over 2} (1 - y^2) x [{4 \over {1-x}} {\hat U}_{,y} 
             + x {\hat U}_{,xy}] \nonumber \\
&&     - {1 \over 4} x^4 ({\hat U}_{,x})^2 (1 - y^2) 
         e^{2((1 - y^2)\hat\gamma)-\beta} 
       - e^{2(\beta-(1 - y^2)\hat\gamma)} \{ 
       - 1  -12 \hat\gamma - 2 y \beta_{,y} \nonumber \\
&&   + (1 - y^2) [
       10 \hat\gamma  
     + 8 y  \hat\gamma_{,y}  
     + 8  \hat\gamma^2 
     + 4 y \hat\gamma \beta_{,y}  
     + \beta_{,yy} 
     + (\beta_{,y})^2  \nonumber \\
&&   - (1 - y^2)^2 (
          8 \hat\gamma^2
        + 2 \hat\gamma_{,y} \beta_{,y} 
        + \hat\gamma_{,yy} 
        + 8 y \hat\gamma\hat\gamma_{,y} )
        + 2 (1 - y^2)^3 (\hat\gamma_{,y})^2 \} .
\label{eq:H-S}
\end{eqnarray}
Note that Eq.~(\ref{eq:xybeta}) has just one radial derivative, and can
be evaluated at the points $(n,i-\frac{1}{2},j)$.  We can discretize it 
as follows
\begin{equation}
\beta_{i,j} =  \beta_{i-1,j} + ({\cal H}_\beta)_{i-\frac{1}{2},j}\, \Delta x .
\label{eq:beta-code}
\end{equation}

Equation (\ref{eq:xyU}), which contains a second radial derivative, is
evaluated at the points $(n,i-1,j)$. Near the origin, where $\hat U
= O(x)$, it becomes a singular differential equation. In order
to integrate it to second order accuracy it is necessary to
apply numerical regularization to the derivatives on the left hand
side. Noting that $\partial/\partial x=4 x^3\partial/\partial x^4$,
Eq.~(\ref{eq:xyU}) becomes
\begin{eqnarray}
2x(1-x)  [x^4 {\hat U}_x]_{,x^4} 
 &+& x^2 [1-(1-x)(\beta_{,x}-(1-y^2)\hat\gamma_{,x})] {\hat U}_{,x} 
\nonumber \\
 &=& e^{2(\beta-(1-y^2)\hat\gamma)} (1-x) {\cal H}_U .
  \label{eq:xyUmod}
\end{eqnarray}
Using the identity 
\begin{equation}
x^4_{i-\frac{1}{2}}-x^4_{i-1-\frac{1}{2}}=2\Delta x\,x_{i-1}(x^2_{i-\frac{1}{2}}+x^2_{i-1-\frac{1}{2}}), 
\end{equation}
we can discretize (\ref{eq:xyUmod}) as
\begin{eqnarray}
& & {(1-x_{i-1}) \over {(x^2_{i-\frac{1}{2}}+x^2_{i-1-\frac{1}{2}})} }
    [x^4_{i-\frac{1}{2}}({\hat U}_{i,j}-{\hat U}_{i-1,j})
    -x^4_{i-1-\frac{1}{2}}({\hat U}_{i-1,j}-{\hat U}_{i-2,j})] \nonumber \\
& + & {1 \over 2} x^2_{i-1} \Delta x [1 - (1-x_{i-1})(\beta_{i,j}-\beta_{i-2,j}
    -(1-y^2_j)(\hat\gamma_{i,j}-\hat\gamma_{i-2,j})){1 \over {\Delta x}}] 
 ({\hat U}_{i,j}-{\hat U}_{i-2,j}) \nonumber \\
& = & (\Delta x)^2 (1-x_{i-1})\: ({\cal H}_U)_{i-1,j} \label{eq:U-disc} .
\end{eqnarray}

The above is a 3-point formula, and it can not be applied at the
points at $x_i=\Delta x$, however we know the asymptotic behavior of
$U$ at the origin, and we can use it to construct a starting algorithm
for these points. We approximate the Bondi equations for $\hat\gamma$
and $\hat U$ by the leading two terms in a power series,
\begin{eqnarray}
\hat\gamma & = & a r^2 + b r^3 \nonumber \\
\hat U     & = & 4 y (a r + \frac{1}{3} b r^2) , \label{eq:u-zero}
\end{eqnarray}
Expanding $\hat\gamma_{,u}$ to the same order, we obtain
for the rate of change of $a$ and $b$
\begin{eqnarray}
    a_{,u} & = & \frac{6}{5} b \nonumber \\
    b_{,u}  & = & 0 . \label{eq:u-zero-dots}
\end{eqnarray}
By fitting a least square polynomial to $\hat\gamma/r^2$ near the origin, we
can read off the coefficients $a$ and $b$ and evaluate $\hat U$ on the
next hypersurface. This approximation is consistent with the global
second order accuracy of the algorithm.
 
As with Eq.~(\ref{eq:xybeta}), Eq.~(\ref{eq:xyS}) can be approximated
at sites $(n,i-\frac{1}{2},j)$ as follows
\begin{eqnarray}
x^2_{i-\frac{1}{2}} {1 \over {\Delta x}} (S_{i,j} - S_{i-1,j})
&+& {x_{i-\frac{1}{2}}\over {1 - x_{i-\frac{1}{2}}}} (S_{i,j} + S_{i-1,j}) 
\nonumber \\
 &=& ({\cal H}_S)_{i-\frac{1}{2},j} .
\label{eq:S-discrete}
\end{eqnarray}

\subsection{The Evolution Equation} \label{sec:evoleqn}

In practice, the corners of the null parallelogram, $P$, $Q$, $R$ and
$S$, cannot be chosen to lie exactly on the grid because the velocity
of light in terms of the compactified coordinate $x$ is not constant
even in flat space. Numerical experimentation suggests that a stable
algorithm with high accuracy results from the choice made in
Fig.~\ref{fig:cell}. 
The essential feature of this placement of the parallelogram with
respect to a coordinate cell is that the sides formed by incoming rays
intersect adjacent u-hypersurfaces at equal but opposite
x-displacements from the neighboring grid points. Solution for the null
geodesics of the metric (\ref{eq:2metric}) to second order accuracy
then gives for the coordinates of the vertices
\begin{eqnarray}
 x_{i-1}- x_P & = & x_R  -  x_{i-1} \nonumber \\
              & = & \Delta u (1 - x_{i-1})^2[1+
                  r_{i-1}(S^{n+1}_{i-2}+S^n_i)/2]/4 \nonumber \\
 x_i- x_Q     & = & x_S  - x_i \nonumber \\
              & = & \Delta u (1 - x_i)^2[1+
                  r_i(S^{n+1}_{i-1}+S^n_{i+1})/2]/4 . \label{eq:box}
\end{eqnarray}
The elementary computational cell consists of the lattice points
$(n,i,j)$ and $(n,i\pm 1,j)$ on the ``old'' hypersurface and the points
$(n+1,i-2,j)$, $(n+1,i-1,j)$ and $(n+1,i,j)$ on the ``new'' hypersurface
(and their nearest neighbors in the angular direction). The marching
algorithm computes the value of the fields at the point $(n+1,i,j)$ in
terms of their predetermined values at the other points in the cell.

The values of $\hat{\psi}$ at the vertices of the parallelogram are
approximated to second order accuracy by linear interpolation between
grid points.  Furthermore, cancellations arise between these four
interpolations so that the evolution equation (\ref{eq:np}) is
satisfied to fourth order accuracy, provided the integral can be
calculated to that accuracy.  This is accomplished by approximating the
integrand by its value at the center of the parallelogram. To second
order accuracy, this gives
\begin{equation}
\int_A du\,dr\,{\cal H} = {\cal H}_c\,\int_A du\,dr 
 = \frac{1}{2} \Delta u\, (r_Q - r_P + r_S - r_R ) {\cal H}_c ,
\label{eq:rhs}
\end{equation}
where the centered value ${\cal H}_c$ can be obtained by averaging
between appropriate points in the cell.
Thus the discretized version of (\ref{eq:np}) is given by 
\begin{eqnarray}
\hat{\psi}^{n+1}_i &=&  {\cal F}(\hat{\psi}^{n+1}_{i-1}, 
     \hat{\psi}^{n+1}_{i-2},      \hat{\psi}^{n}_{i+1},
     \hat{\psi}^{n}_{i},          \hat{\psi}^{n}_{i-1}) \nonumber \\
&& + \frac{1}{2} \Delta u\, (r_Q - r_P + r_S - r_R ) {\cal H}_c
\label{eq:disc_evo}
\end{eqnarray}
where ${\cal F}$ is a linear function of $\psi$ and the $j$ index
has been suppressed. 

Consequently, it is possible to move through the interior of the
lattice, computing $\hat{\psi}^{n+1}_i$ explicitly by an orderly radial
march.  This is achieved by starting at the origin at time $u^{n+1}$.
Field values vanish there.  Next, proceed outward one radial step using
the boundary conditions (discussed below). Then step outward to the
next interior radial point using (\ref{eq:disc_evo}), iterating this
process throughout the interior and for all angles.  This updates all
field values stretching to scri along the new null cone at $u^{n+1}$,
thus completing one evolutionary time step. 

The above scheme is sufficient for accurate evolution in a neighborhood
of the origin. Global evolution, including the points at $x=1$,
requires careful manipulation of (\ref{eq:disc_evo}) to avoid problems
from the fact that $\psi \sim r K $ at scri.  Thus the direct use of
this formula is not possible for the point {\em at} scri, while points
{\em near} scri would suffer serious loss of accuracy.  We renormalize
(\ref{eq:disc_evo}) in the following way.  First, we introduce the
quantity $\phi = \psi (1 - x) $. Near scri $\phi$ has the desired
finite behavior, while near the origin it leaves unchanged the constant
coefficient form of the evolution equation, thus preserving the
stability properties.  With this substitution and with the use of
(\ref{eq:rhs}), the evolution equation (\ref{eq:np}) becomes
\begin{eqnarray}
     \phi_Q &=&  {1\over4} \Delta u\, x_{Q} {\cal H}_{c}
   + \frac{(1 - x_{Q})}{(1 - x_{P})} 
     (\phi_P - {1\over4} \Delta u\, x_{P} {\cal H}_{c}) \nonumber \\
&&  + \frac{(1 - x_{Q})}{(1 - x_{S})} 
     (\phi_S + {1\over4} \Delta u\, x_{S} {\cal H}_{c}) 
    - \frac{(1 - x_{Q})}{(1 - x_{R})} 
     (\psi_R + {1\over4} \Delta u\, x_{R} {\cal H}_{c})          
\label{eq:phinp}
\end{eqnarray} 
Now all terms have finite asymptotic value. The coefficient $(1 -
x_{Q})/(1- x_{S})$ has $0/0$ behavior at scri but approaches the limit
$1$. Further refinement is possible with the use of the explicit
second order approximation for the characteristics (\ref{eq:box}) which
leads to the approximation
\begin{equation}
\frac{(1 - x_{Q})}{(1- x_{S})} = 1 + 2 \frac{\delta}{1-\delta}
\end{equation}
where
\begin{equation}
\delta = {1\over4} \Delta u (1 + x_i(S^{n+1}_{i-1}+S^n_{i+1})/2)
\end{equation}
The final result is that the equation (\ref{eq:phinp}) propagates $\phi$
radially outward one cell with an error of fourth order in grid size.
This is valid for all interior points and the point at scri.
The error in each cell compounds to a third order error on each null
cone and a second order global error after evolving for a given
physical time. Second order global accuracy is indeed confirmed by the
convergence tests described in Sec.~\ref{sec:tests}.

We mentioned that a modified form of the basic grid cell
Eq.~(\ref{eq:box}) is used at the origin. This is necessary since the
incoming characteristic through the points $P$ and $R$ can not be
centered at $x=0$.  The corners of the modified cell are given by
\begin{eqnarray}
 x_P & = & 0 \nonumber \\
 x_R & = & \frac{1}{2} \Delta u \nonumber \\
 x_1 - x_Q    & = & x_S  - x_1 = \frac{1}{4} \Delta u (1 - x_i)^2 . \label{eq:box-zero}
\end{eqnarray}
Only the linear terms of ${\cal H}$ are kept while evolving the first
point, i.e. for $x_1=\Delta x$. This reduces  equation (\ref{eq:np}) to
\begin{equation}
    \hat{\psi}_Q = \hat{\psi}_S - \hat{\psi}_R
     - {1 \over 4} \int_A du\, dr \frac{1}{r} (r^2 \hat U_{,y})_{,r}.
\end{equation}
Using the expansion  (\ref{eq:u-zero}) for $U$ near the origin, the
integral simplifies further to
\begin{equation}
     \int_A du\, dr (3 \, a\, r + \frac{12}{5} \, b\, r^2).
\end{equation}
The integrand is now evaluated to second order accuracy at
$u^{n+\frac{1}{2}} = u^n + \frac{\Delta u}{2}$ using
$$a^{n+\frac{1}{2}} = a^n + \frac{6}{5} b^n \frac{\Delta u}{2}$$ 
and $b^{n+\frac{1}{2}} = b^n$.  Keeping higher order
terms would not improve the global convergence rate of the code.

\section{Code tests} \label{sec:tests}

\subsection{Testbeds}

As we have shown in Sec.~\ref{sec:linear}, linearized solutions of the
Bondi equations can be generated by solutions of the scalar wave
equation, thus supplying a complete set of test beds for the very weak
regime.  For the nonlinear case, exact boost and rotation symmetric
solutions~\cite{bicschmidt} of the Bondi initial hypersurface equations have also
been found~\cite{bic}.  They have been used to check the radial
integrations leading from $\gamma$ to $\beta$, $U$ and $V$ but they do
not provide a test of the evolution algorithm. However, in the course
of this work, we have found that one of these initial data sets is in
fact preserved under time evolution and is an exact static solution of
the nonlinear vacuum equations.

This solution provides an important test bed for null cone evolution
codes. Except for spherically symmetric cases, it is the only known
solution of Einstein's equation which can be expressed explicitly in
null cone coordinates with no singularity at the vertex. It
has the form
\begin{eqnarray}
2\>e^{\gamma}&=&1+\Sigma \nonumber \\
e^{2\beta}&=&{(1+\Sigma)^2\over{4\Sigma}}\nonumber \\
U&=&-{{a^{2}\rho\>\sqrt{r^{2}-\rho^{2}}}\over{r\>\Sigma}}\nonumber \\
V&=&{{r\>\left(2\>a^{2}\rho^{2}-a^{2}\>r^{2}+1\right)}\over{\Sigma}}  
\label{eq:simple}
\end{eqnarray} 
where $\rho = r \sin\theta$, $\Sigma=\sqrt{1+a^2\rho^2}$ and $a$ is a
free scale parameter. It is remarkable that the null data $\gamma$ is
time independent under evolution, as can be verified by inserting the
above expressions in (\ref{eq:gammaev}).
Because this solution is static, as well as boost and rotation
symmetric, the commutator between the boost and time translation
symmetries implies that it has an additional translation symmetry in
the boost direction. Thus the solution falls into several overlapping
and widely studied classes, including the static cylindrically
symmetric spacetimes and the Weyl static axisymmetric spacetimes.
However, this solution, which we call SIMPLE, has not previously been
singled out, apparently because it cannot easily be identified in the
traditional coordinates used for studying static solutions. Because of
its cylindrical symmetry, it is clear that this solution is not
asymptotically flat but it can be used to construct an asymptotically
flat, nonsingular solution by smoothly pasting asymptotically flat null
data to it outside some radius $R$.  The resulting solution will be
static and given by (\ref{eq:simple}) in the domain of dependence
interior to $R$. Numerical solutions generated by this technique are
used in the code calibration tests presented below.

In addition, global energy conservation provides an important test bed.
The Bondi mass loss formula is not one of the equations used in the
evolution algorithm but follows from those equations as a consequence
of a global integration of the Bianchi identities.

\subsection{Convergence and Stability}

We have tested the algorithm to be second order accurate and stable,
subject to the CFL condition, throughout the regime in which caustics
and horizons do not form. In Sec.~\ref{sec:linear}, we showed how the
the linearized Bondi equations may be reduced to the scalar wave
equation by local operations. For very weak data, the nonlinear
equations approximate the linear equations so that we would expect the
global stability of the nonlinear algorithm to be related to the CFL
condition for the scalar wave algorithm.  Near the origin, stability
checks show that the time step is limited by (\ref{eq:cfl}) with $K=8$,
which is twice the limit found for the scalar wave algorithm. This
factor of two apparently arises from the use of a staggered grid in the
gravitational case, which effectively doubles the value of $r$ at which
the main algorithm takes over from the start up algorithm at the
origin. This gives some reassurance that the scalar wave algorithm has
been optimally adapted to the Bondi equations.

By construction, the $u$-direction is timelike at the origin where it
coincides with the worldline traced out by the vertex of the outgoing
null cone. But even for weak fields, the $u$-direction becomes
spacelike at large distances along a typical outgoing ray. This can be
seen from the metric coefficient $g_{uu}=(V/r)e^{2 \beta}-U^2 r^2 e^{2
\gamma}$ which at large $r$ becomes dominated by the the asymptotic
behavior $U=L+O(1/r)$. Geometrically, this reflects the property that
scri is itself a null hypersurface so that all internal
directions are spacelike, except for the null generator. For a flat
space time, the $u$-direction picked out at the origin corresponds to
the null direction at scri but it becomes asymptotically spacelike
under the slightest deviation from spherical symmetry.

\begin{figure}
\centerline{\psfig{figure=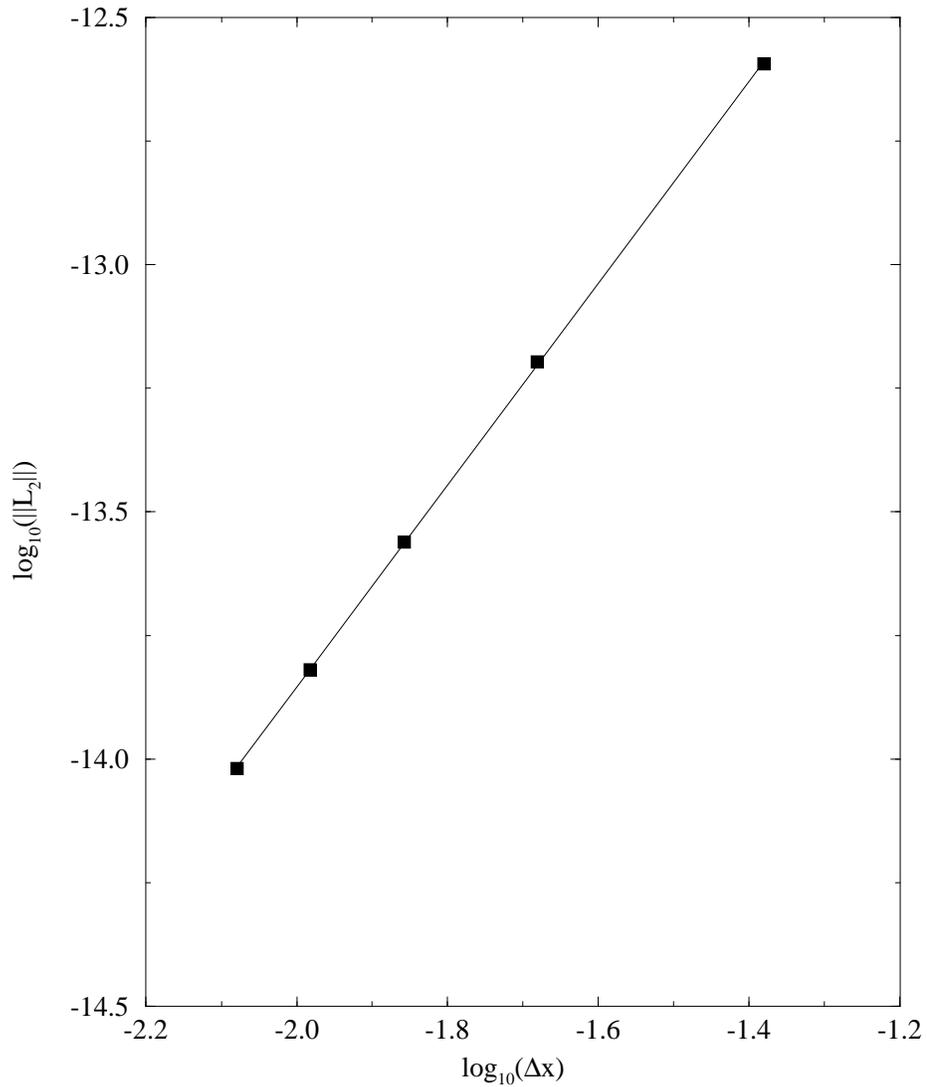,height=7.765in,width=6.0in}}
\caption{The error, as measured by the $L_{2}$ norm, of a numerical
solution with higher harmonics ($\ell=6$). The computation is made on
grids of size $N_x$ equal to 24, 48, 72, 96 and 120, while keeping
$N_x=3N_y$. }
\label{fig:l6}
\end{figure}

By choosing initial data of very small amplitude ($|\hat\gamma|\approx
10^{-9}$), we have performed convergence tests of the numerical
solutions against the solutions of the linearized equations. The
linearized solutions (\ref{eq:ltest}) were given as initial data at
$u=0$ and we compared the numerically evolved solutions to the
linearized solutions at a central time of $u=0.5$. We observed that for
the low angular momentum solutions ($\ell=$2, 3, 4) the code is
superaccurate, i.e. the solutions converge to the exact result at a
rate faster than second order in the grid size.  This is to be expected
since for these solutions the hatted variables used in the code exhibit
angular dependence that is at most quadratic in $y$, so that the second
order accurate $y$-derivatives are calculated exactly. The error, as
measured by the $L_{2}$ norm, of a numerical solution with higher
harmonics ($\ell=6$) is graphed in Fig.~\ref{fig:l6}. The slope of the
graph gives a convergence rate of $2.04\pm 0.01$ with respect to grid
size. This result is insensitive to the particular norm used, i.e. we
also verified second order convergence in the $L_{1}$ and $L_{\infty}$
norms.

\begin{figure}
\centerline{\psfig{figure=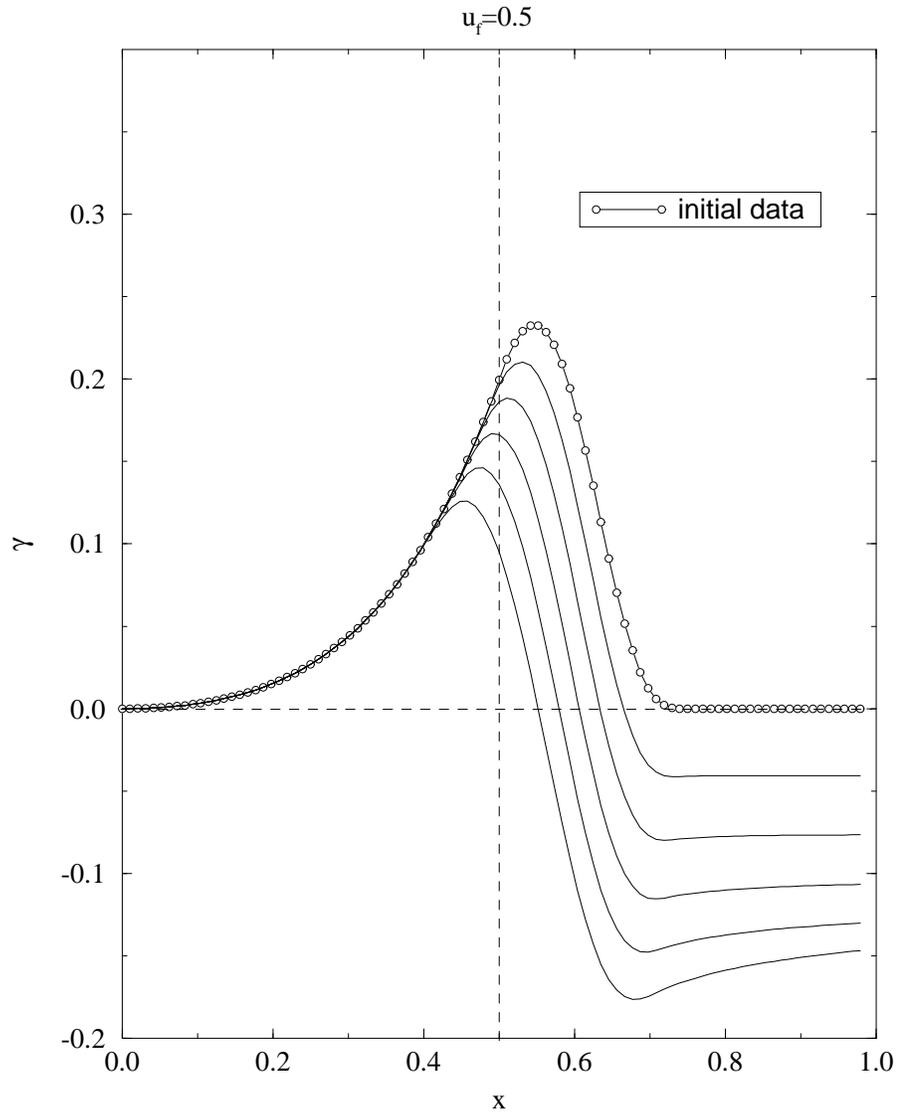,height=7.765in,width=6.0in}}
\caption{Evolution of initial data given by SIMPLE in the interior region,
and patched smoothly to an asymptotically flat exterior. The static
interior is preserved to graphical accuracy.}
\label{fig:simple}
\end{figure}

\begin{figure}
\centerline{\psfig{figure=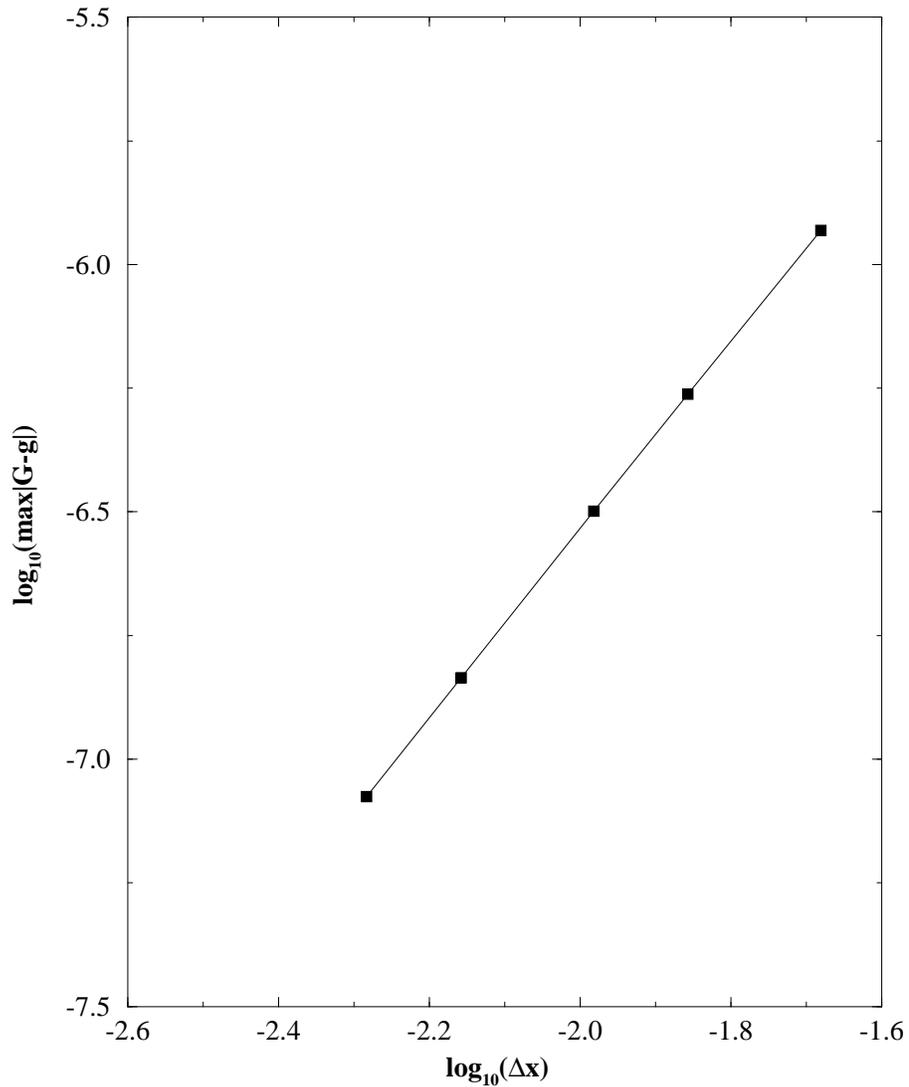,height=7.765in,width=6.0in}}
\caption{The error in the evolution of the initial data of
Fig.~\protect\ref{fig:simple} up to $u=0.25$, as measured by the
$L_{\infty}$ norm. The error is computed on grids of size $N_y$ equal
to 16, 24, 32, 48 and 64, while keeping $N_x=3N_y$. The convergence
rate is $1.92$, in good agreement with the theoretically expectation of
second order accuracy.}
\label{fig:convergence}
\end{figure}

Second order convergence has also been checked against the exact static
solution SIMPLE. Since this solution is not asymptotically flat, we
match the initial data smoothly to asymptotically flat data for a
nonstatic exterior.  Consideration of the domain of dependence implies
that the matching boundary propagates along an ingoing null
hypersurface. Thus, we can obtain a reliable measure of how accurately
the evolution preserves the static interior, provided we restrict the
calculation of the error norm to a region not yet influenced by the
exterior nonstatic data. We matched one such static solution in the
interior ($x\le 0.5$) to smooth exterior data with compact support.  We
calculated the $L_{\infty}$ norm for the region $x\le 0.4$ at time
u=0.25, and considered the dependence of the error on the grid size.
The preservation of the static interior to graphical accuracy is shown
in Fig.~\ref{fig:simple}, while the second order convergence of the
error is demonstrated in Fig.~\ref{fig:convergence}.  In addition, for
a wide variety of initial data having unknown analytic solution, we
have verified that the numerical solution converges to second order in
the sense of Cauchy convergence.

\begin{figure}
\centerline{\vbox{\hbox{\psfig{figure=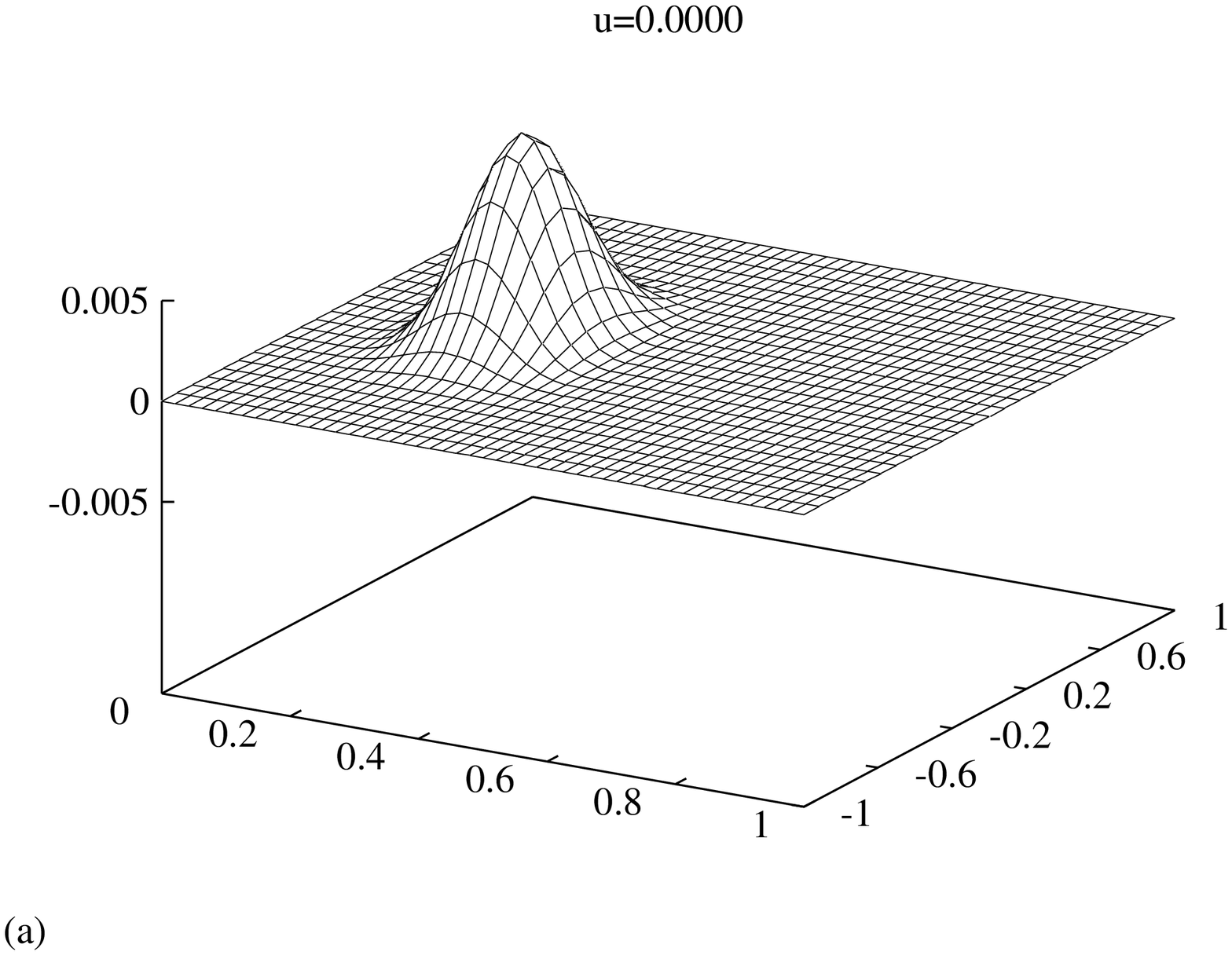,height=2.8in,width=3.75in}}
                   \hbox{\psfig{figure=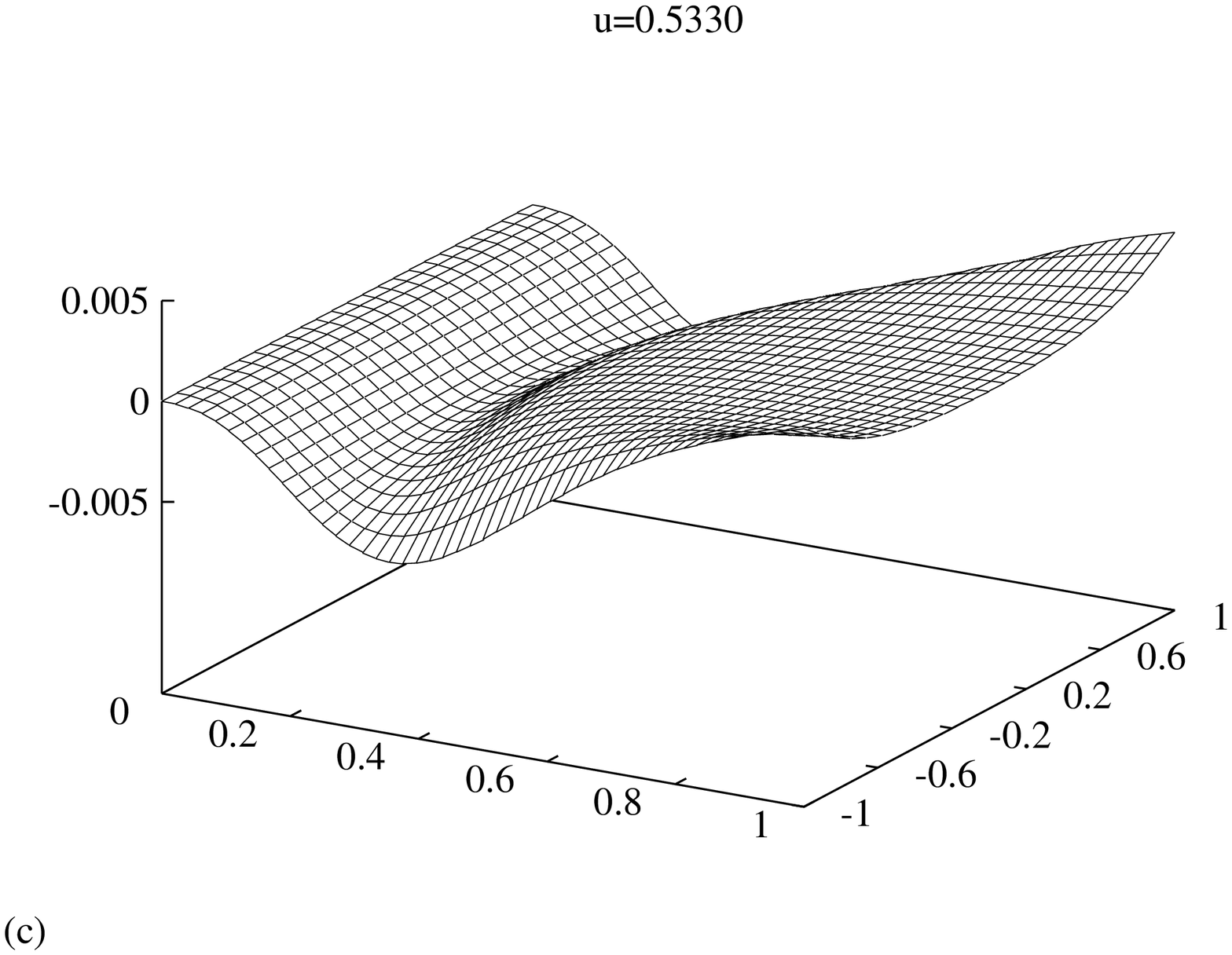,height=2.8in,width=3.75in}}
                   \hbox{\psfig{figure=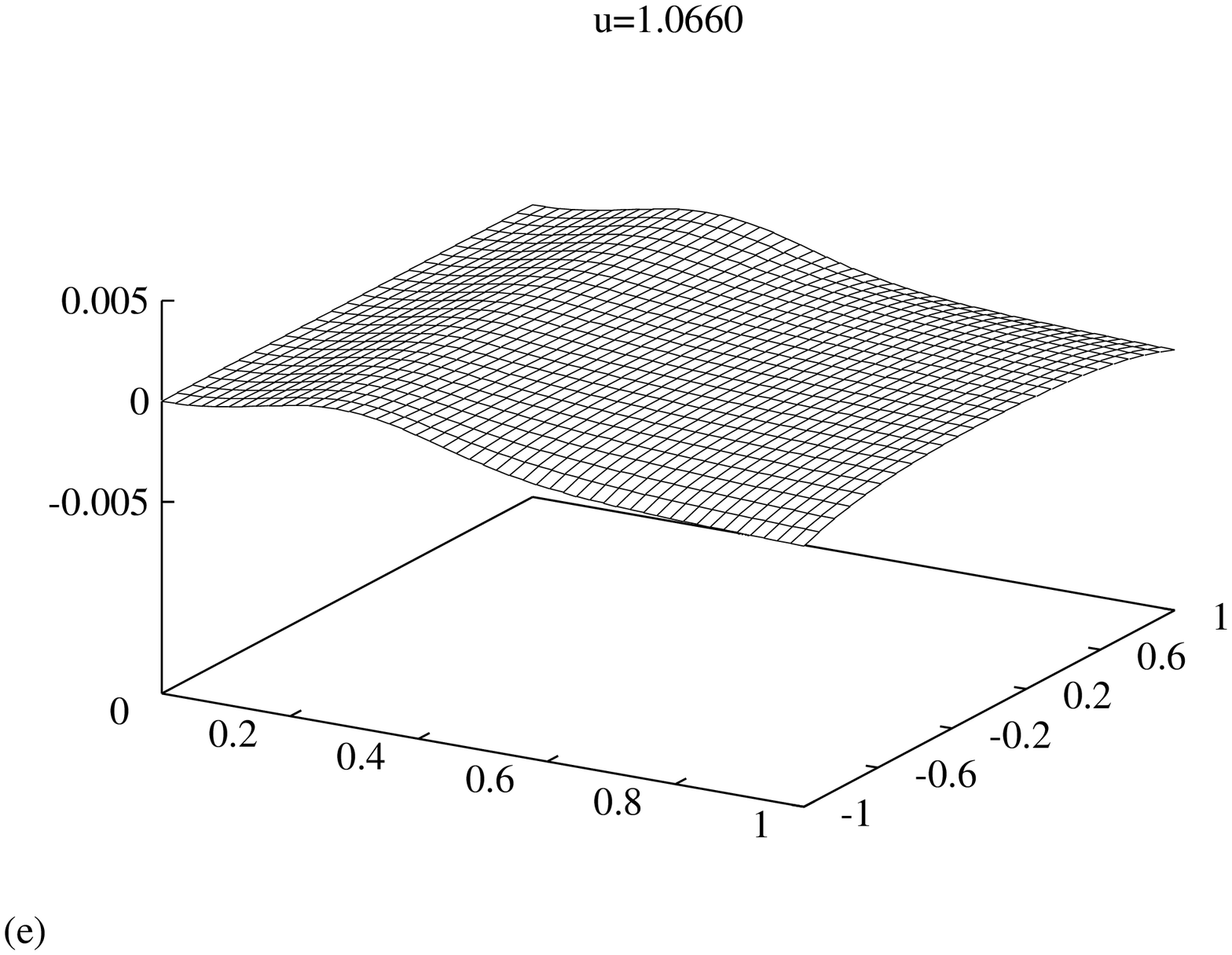,height=2.8in,width=3.75in}}}
             \vbox{\hbox{\psfig{figure=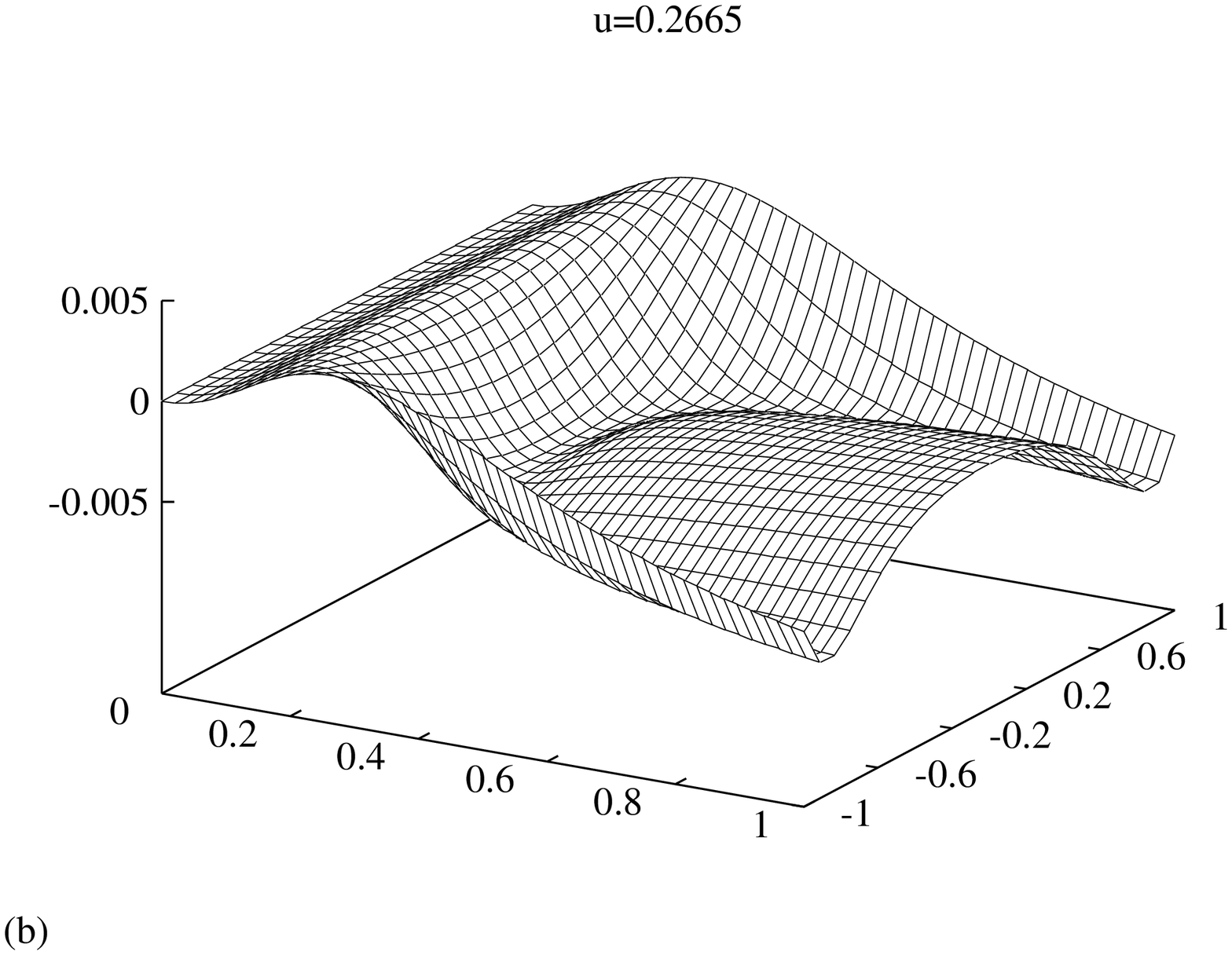,height=2.8in,width=3.75in}}
                   \hbox{\psfig{figure=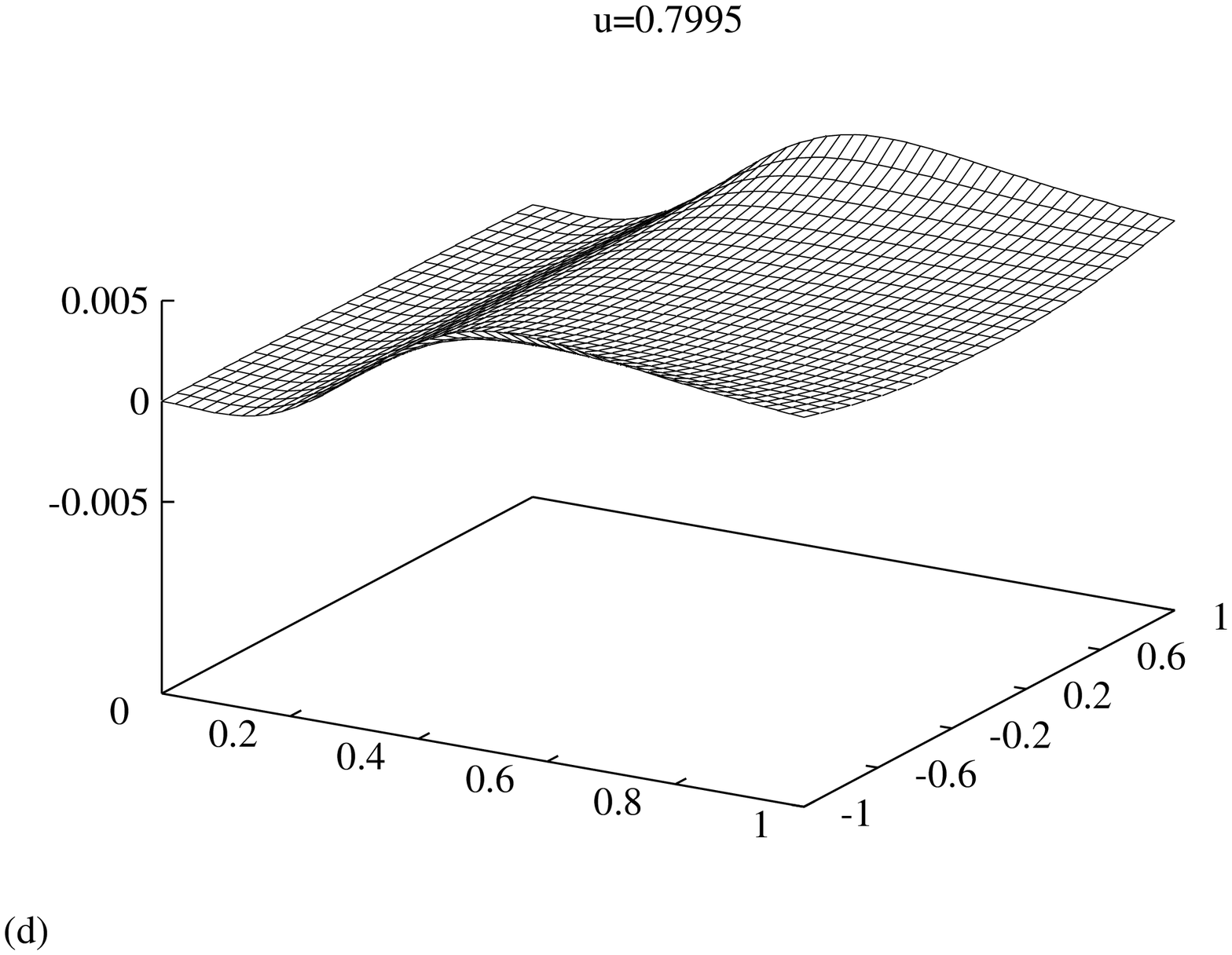,height=2.8in,width=3.75in}}
                   \hbox{\psfig{figure=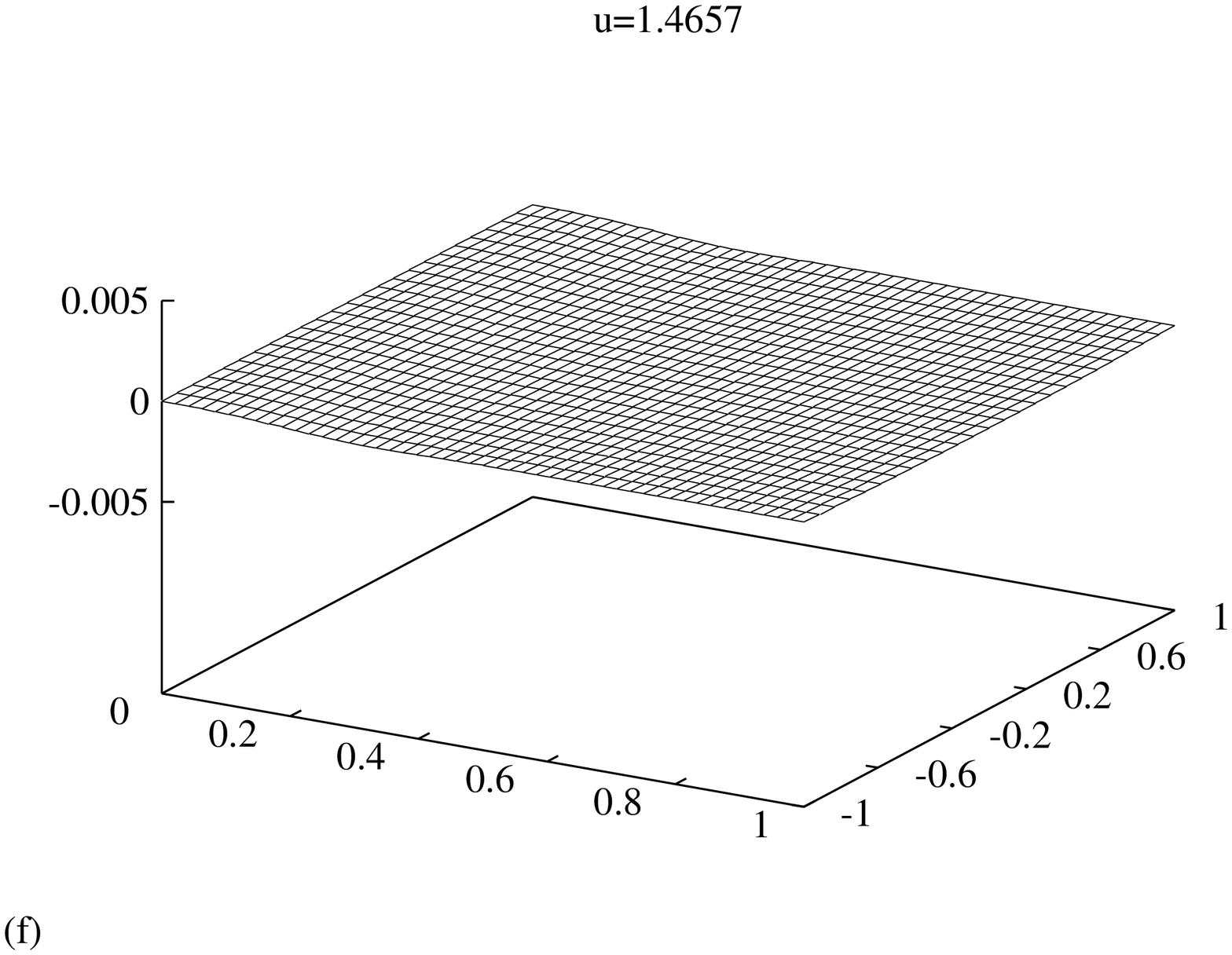,height=2.8in,width=3.75in}}}
             }
\caption{A sequence of time slices of the numerical evolution of
initial data of compact support. Note all the angular structure that
arises at about $u=0.25$, which later decays. At $u\approx 1.5$ the
amplitude of the field has decayed below those values which can be
observed in the figure.}
\label{fig:seq}
\end{figure}

Stability of the code in the low to medium amplitude regime has been
verified experimentally by running arbitrary initial data until it
radiates away to scri. At higher amplitudes, it is expected that
physical singularities will arise, but we have not yet explored this
regime.  Figure~\ref{fig:seq} shows a sequence of time slices of the
numerical evolution of some arbitrary initial data of compact support.
Note the rich angular structure that arises at $u\approx 0.25$ and then
dissipates. At $u\approx 1.5$ the amplitude of the field is
sufficiently small so that it appears to be zero in the figure. It
continues to decay at later times.

\subsection{Energy Conservation}

The Bondi mass loss formula relates the gravitational radiation power
to the square of the news function. It follows from the equations used
in the algorithm as a consequence of a global integration of the
Bianchi identities. Thus it not only furnishes a valuable tool for
physical interpretation but it also provides a very important
calibration of numerical accuracy and consistency.

Historically, numerical calculations of the Bondi mass $M_B$ have been
frustrated by technical difficulties arising from the necessity to pick
off nonleading terms in an asymptotic expansion about infinity. For
example, the mass aspect ${\cal M}$ must be picked off in the the asymptotic
expansion (\ref{eq:Vasym}) for $V$. This is similar to the experimental
task of determining the mass of an object by measuring its far field.
In the non-radiative case it can be accomplished by measuring gravity
gradients, but otherwise this approach can be swamped by radiation
fields. In the computational problem, further complications arise from
gauge terms which dominate asymptotically even over the radiation
terms.  We have recently developed a second order accurate algorithm
for calculating the Bondi mass ~\cite{mbondi}.  It avoids the above
problems through the use of Penrose compactification and the
introduction of renormalized variables in which Bondi's mass aspect
appears as the leading asymptotic term.  The Bondi mass algorithm
depends only upon fields on a single null hypersurface. It has been
incorporated into the present evolution code to calculate the mass at
any given retarded time.

In the present formalism, the news function $N$ is given by~\cite{IWW}
\begin{equation}
2\,e^{2H} N = 2\,c_{,u}
+{ {(\sin \theta\,c^2\,L)_{,\theta}} \over {\sin\theta\,c} }
+e^{-2K}\omega\sin\theta \left[ {(e^{2H} \omega)_{,\theta} \over {\sin\theta\,\omega^2}} \right] _{,\theta}.
\end{equation}
Here $\omega$ is the conformal factor relating the asymptotic
2-geometry to the unit sphere geometry of a Bondi frame, i.e.
\begin{equation}
   e^{2K} d\theta^2 +\sin^2\theta e^{-2K} d \phi^2 =
	{\omega^{-2}}( d\theta_B^2+\sin^2\theta_B \, d \phi_B^2 ),
\end{equation}
where $\theta_B$ and $\phi_B=\phi$ are Bondi spherical coordinates.
Calculation of $\omega$ complicates the calculation of the news
function. The simplest approach is to set $y=-\cos\theta$ and 
$y_B=-\cos\theta_B$.  Then
\begin{equation}
\omega^2 = {dy_B \over dy} , \label{eq:omegay}
\end{equation}
where
\begin{equation}
 y_B = \tanh\left[\int_0^y\,{dy \over {1-y^2}}\,e^{2K}\right] . \label{eq:yb}
\end{equation}
This gives
\begin{equation}
 \omega = {2\,e^K \over {(1+y)e^\Delta +(1-y)e^{-\Delta}} }
\end{equation}
where
\begin{equation}
 \Delta = \int_0^y dy\,{{e^{2K}-1} \over {1-y^2}}. \label{eq:delta}
\end{equation}
In order to prepare this integral in an explicitly regular form for computation we
introduce an auxiliary parameter $\alpha$ and rewrite (\ref{eq:delta}) 
as the double integral
\begin{equation}
 \Delta = 2\int_0^y dy\int_0^1 d\alpha \, e^{2\alpha K}\hat{K},  
    \label{eq:alphdelta}
\end{equation}
where $\hat{K}=K/(1-y^2)$ is regular at the poles.  It is then
straightforward to obtain a second order accurate finite difference
formula for the news function.

\begin{figure}
\centerline{\psfig{figure=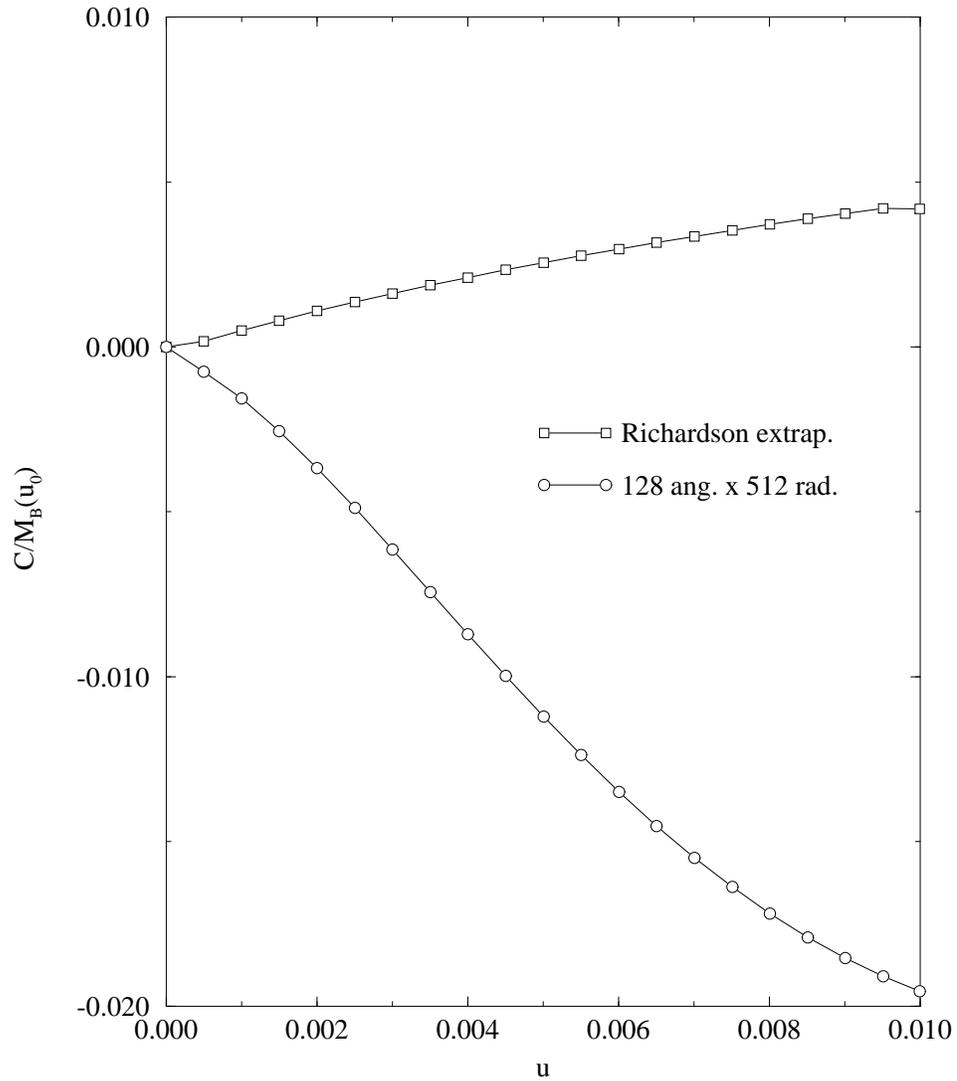,height=7.765in,width=6.0in}}
\caption{Graph of the relative error in $C$ calculated up to
$u=0.01$ for a numerical evolution of the data
(\protect\ref{eq:compact}), on a grid of 512 radial $\times$ 128
angular points. The circles show the error as calculated from the computed
values of the mass, while the squares show the error after using Richardson
extrapolation, based on the known convergence rate of the algorithm.}
\label{fig:masscons}
\end{figure}

The Bondi formula for energy conservation between central times
$u_0$ and $u$ takes the form $C$ =0, where
\begin{equation}
 C=M_B(u)-M_B(u_0)+
    {1\over 2}\int_{-1}^1 dy\int_{u_0}^{u} du\, e^{2H} \omega^{-1} N^2.
\end{equation}
Figure~\ref{fig:masscons} graphs $C$ relative to the initial
mass $M_B(u_0)$ for a numerical evolution of the polynomial data
\begin{equation}
	\hat\gamma = \lambda 
{ \left[{(x - x_1)(x - x_2)(y^2 - y_0^2)}\right]^6
\over {\left[ (x_1 - x_2) y_0 \right]^{12}} }
\label{eq:compact}
\end{equation}
with compact support in the domain $(x_1\le x\le x_2) \times (-y_0 \le
y \le y_0)$ and amplitude parameter $\lambda$. For the graph, we have
chosen $\lambda=0.3$, $x_1=0.1$, $x_2=0.5$ and $y_0=0.5$, and evolved
the numerical solution up to $u=0.01$, on a grid of 512 radial $\times$
128 angular points.
The bulk of the error occurs in the calculation of the
Bondi mass, whose accuracy is more sensitive to grid size than the
accuracy of either the news function or the evolution code.
Most of this error may be removed by using Richardson extrapolation to
take advantage of the known second order accuracy of the Bondi mass.
For example, if $F_n(x_i)$ is a second order accurate finite difference
approximation to the function $f(x)$ on a grid of $n$ points, then
$(4F_n-F_{n/2})/3$ approximates $f$ to third order and in fact to
fourth order if odd orders are absent in the approximation. This
absence of odd orders indeed holds for the Bondi mass because all
derivatives, interpolations and integrals are centered. 
Thus, introduction of subgrids obtained by subsampling, leads to a
fourth order expression for $M_B$, with the corresponding relative
error in energy conservation also graphed in Fig.~\ref{fig:masscons}.
In this way, energy conservation is attained to 0.4\% accuracy.
Note that only a single evolution on a fixed grid is
necessary here because Richardson extrapolation is applied when
calculating the Bondi mass.  For the purposes of the graph, we have
done this for each time that $C$ is plotted, but to check energy
conservation, it suffices to do it only at the initial and final
times.  Figure~\ref{fig:masscons} serves as a rewarding testament to
the virtues of a code with known convergence rates.
 
\section{Conclusion} \label{sec:discussion}

We have constructed a second order accurate evolution algorithm for the
null cone initial value problem for axisymmetric vacuum spacetimes.
Energy conservation is maintained to second order accuracy.  Extensive
tests of the algorithm establish that it is globally valid in the
regime where horizons and caustics do not develop. This generates a
large complement of highly accurate numerical solutions for the class
of asymptotically flat, axisymmetric vacuum spacetimes, for which no
analytic solutions are known. All results of numerical evolutions in
this regime are consistent with the theorem of Christodoulou and
Klainerman~\cite{XKlain} that weak initial data evolve asymptotically
to Minkowski space at late time. The code is now being tested in the
strong field regime for application to the study of black hole
formation.

\acknowledgements

We benefited from research support from the National Science
Foundation under NSF Grant PHY92-08349 and from computer time made
available by the Pittsburgh Supercomputing Center.

\appendix\section*{}

We sketch here the von Neumann stability analysis of the algorithm for
the linearized Bondi equations. The analysis is based up freezing the
explicit functions of $r$ and $y$ that appear in the equations, so that
it is only valid locally for grid sizes satisfying $\Delta r << r$ and
$\Delta y <<1$. However, as is usually the case, the results are
quite indicative of the stability of the actual global behavior of
the code. 
 
Starting with the hatted code variables introduced in Sec.~\ref{sec:fde}
and setting $\Gamma=r^2 \hat U$ and $G=r\hat\gamma$,
the linearized Bondi equations (\ref{eq:lu}) and (\ref{eq:lgammaev})
take the form
\begin{equation}
   r^2 \Gamma_{,rr}-2\Gamma=2[4y-(1-y^2)\partial_y](rG_{,r}-G)
  \label{eq:gamma}
\end{equation}
and
\begin{equation}
        2G_{,ur}-G_{,rr}=-(1/2r)\Gamma_{,ry}.
\label{eq:bigg}
\end{equation}
Freezing the explicit factors of $r$ and $y$ at $r=R$ and $y=Y$,
introducing the Fourier modes $G=e^{su}e^{ikr}e^{ily}$ (with real $k$
and $l$) and setting $\Gamma=AG$, these equations imply
\begin{equation}
  A=2(1-ikR)[4Y-(1-Y^2)il]/(2+R^2 k^2)
\end{equation}
and
\begin{equation}
  4is=-2k+Al/R.
\end{equation}
For stable modes, $Re(s)\ge 0$. This requires that the $lIm(A)\le 0$
which will be satisfied unless $Ykl<0$. In the latter case, unstable
solutions exist to the  PDE's obtained by freezing the coefficients in
the linearized equations (\ref{eq:gamma}) and (\ref{eq:bigg}). The
linearized equations themselves do not have unstable modes but they
arise in the frozen coefficient formalism from dropping the boundary
condition of spherical topology on  the $y$-dependence. For a global
solution, $G$ should not have periodic dependence on $y$ but instead be
decomposed into spin-weight 2 harmonics, in which case instabilities
would not arise in the above analysis. Thus these unstable modes of the
frozen PDE are artificial and should be discarded by requiring $Ykl\ge
0$ when analyzing the stability of the corresponding FDE.

Consider now the FDE obtained by putting $G$ on the grid points $r_I$
and $\Gamma$ on the staggered points $r_{I+1/2}$, while using the same
angular grid $y_J$ and time grid $u_N$. Let $P$, $Q$, $R$ and $S$ be
the corner points of the null parallelogram algorithm, placed so that
$P$ and $Q$ are at level $N+1$, $R$ and $S$ are at level $N$, and so
that the line $PR$ is centered about $r_I$ and $QS$ is centered about
$r_{I+1}$.  For simplicity, we display the analysis at the equator
where $Y=0$.  Then, using linear interpolation and centered derivatives
and integrals, the null parallelogram algorithm for the frozen version
of the linearized equations leads to the FDE's
\begin{eqnarray}
  & (R/\Delta r)^2(\Gamma_{I+3/2}-2\Gamma_{I+1/2}+\Gamma_{I-1/2})
        -(\Gamma_{I+3/2}+\Gamma_{I-1/2}) \nonumber \\
  & =-\delta_y[2(R/\Delta r)(G_{I+1}-G_{I})-(G_{I+1}+G_{I})]
\label{eq:fgamma}
\end{eqnarray}
(all at the same time level) and
\begin{eqnarray}
 & & G_{I+1}^{N+1}-G_I^{N+1}-G_{I+1}^N+G_I^N \nonumber \\
 &+& (\Delta u/4\Delta r)
  (-G_{I+1}^{N+1}+2G_I^{N+1}-G_{I-1}^{N+1}-G_{I+2}^N+2G_{I+1}^N-G_I^N)
  \nonumber \\
 &=&-(\Delta u/8R)\delta_y(\Gamma_{I+1/2}^{N+1}-\Gamma_{I-1/2}^{N+1}
    +\Gamma_{I+3/2}^N-\Gamma_{I+1/2}^N),
        \label{eq:fbigg}
\end{eqnarray}
where $\delta_y$ represents a centered first derivative. Again setting
$\Gamma=AG$ and introducing the discretized Fourier mode $G=e^{sN\Delta
u}e^{ikI\Delta r}e^{ilJ\Delta y}$, we have $\delta_y=i\sin(l\Delta
y)/\Delta y$ and (\ref{eq:fgamma}) and (\ref{eq:fbigg}) reduce to
\begin{equation}
   A[(R/\Delta r)^2(1-\cos\alpha)+\cos\alpha]
    =-L[(2R/\Delta r)\sin(\alpha/2)+i\cos(\alpha/2)]
  \label{eq:ffgamma}
\end{equation}
and 
\begin{equation}
       e^{s\Delta u}=-e^{i\alpha}(C^*-AD)/(C-AD),
       \label{eq:stab}
\end{equation}
where $L=\sin(l\Delta y)/\Delta y$, $\alpha =k\Delta r$,
$C=ie^{i\alpha/2}\sin(\alpha/2)+(\Delta u/4\Delta r)(1-\cos\alpha)$
and $D=(L\Delta u/8R)\sin(\alpha/2)$. The stability condition that
$Re(s)\le 0$ then reduces to $Re[CD(A-A^*)]\ge 0$ which is equivalent
to $1+\cos\alpha[1-(\Delta r/R)^2] \ge 0$. Thus this stability condition
is automatically satisfied and poses no constraint on the algorithm.

The corresponding analysis at the poles $Y=\pm 1$ again leads to
(\ref{eq:stab}), where now
\begin{equation}
   A[(R/\Delta r)^2(1-\cos\alpha)+\cos\alpha]
    =-4Y[2i(R/\Delta r)\sin(\alpha/2)-\cos(\alpha/2)].
  \label{eq:pffgamma}
\end{equation}
The stability condition $Re[CD(A-A^*)]\ge 0$ is satisfied provided
$Ykl\ge 0$, which rules out the artificially unstable solutions of the
frozen PDE discussed above.

As a result, local stability analysis places no constraints on the
algorithm. This may seem surprising because not even the analogue of a
CFL condition on the time step arises but it can be understood in the
following vein. The local structure of the code is implicit, since it
involves 3 points at the upper time level.  Implicit algorithms do not
necessarily lead to a CFL condition. However, the algorithm is globally
explicit in the way that evolution proceeds by an outward radial
march from the origin. It is this feature that necessitates a CFL
condition in order to make the numerical and physical domains of
dependence consistent.


\newpage

\end{document}